\documentstyle[aps]{revtex}

\begin{document}

\title{Multifractal behavior of linear polymers in disordered media}

\author{Anke~Ordemann$^{1,2}$, Markus~Porto$^{3}$,
        H.~Eduardo~Roman$^{4}$, Shlomo~Havlin$^{2,1}$, and~Armin~Bunde$^{1}$}

\address{$^1$Institut~f\"ur~Theoretische~Physik~III,
         Justus-Liebig-Universit\"at~Giessen,
         Heinrich-Buff-Ring~16,
         35392~Giessen, Germany}

\address{$^2$Minerva~Center~and~Department~of~Physics,
         Bar-Ilan~University,
         52900~Ramat-Gan, Israel}

\address{$^3$School~of~Chemistry,
         Tel~Aviv~University, 
	 69978~Tel~Aviv, Israel}

\address{$^4$I.N.F.N., Sezione~di~Milano,
         Via~Celoria~16, 20133~Milano, Italy}

\date{received \today}
\maketitle

\begin{abstract}
The scaling behavior of linear polymers in disordered media modelled by
self-avoiding random walks ({\it SAW}s) on the backbone of two- and
three-dimensional percolation clusters at their critical concentrations $p_{\rm
c}$ is studied. All possible {\it SAW} configurations of $N$ steps on a single
backbone configuration are enumerated exactly. We find that the moments of
order $q$ of the total number of {\it SAW}s obtained by averaging over many
backbone configurations display multifractal behavior, i.e.\ different moments
are dominated by different subsets of the backbone. This leads to generalized
coordination numbers $\mu_q$ and enhancement exponents $\gamma_q$, which depend
on $q$. Our numerical results suggest that the relation $\mu_1 = p_{\rm c} \mu$
between the first moment $\mu_1$ and its regular lattice counterpart $\mu$ is
valid.
\end{abstract}

\bigskip

\newpage

\section{Introduction}

The question of how linear polymers behave in a disordered medium has attracted
much attention in recent years. The problem is not only interesting from a
theoretical point of view, but may also be relevant for understanding transport
properties of polymeric chains in porous media, such as in enhanced oil
recovery, gel electrophoresis, gel permeation chromatography, etc.\
\cite{Doi/Edwards:1986,Dullien:1979,Andrews:1986,Baumgaertner:1995}. In this
context, it is useful to learn about the static or conformational properties of
linear chains, modelled by self-avoiding walks ({\it SAW}s), in the presence of
quenched disorder, e.g.\ how the surrounding structural disorder influences
their spatial configuration. As a quite general model of a random medium,
percolation~\cite{Flory:1941,Stauffer/Aharony:1992,Sahimi:1994,%
Bunde/Havlin:1996} may be considered as the paradigm for a broad class of
disordered systems and has therefore been mostly used so far.

We are interested in how the statistical behavior of {\it SAW}s  on percolation
clusters at criticality ($p = p_{\mathrm{c}}$) differs from their behavior on
regular lattices. While the values of the exponents for {\it SAW}s on regular
lattices are well
established~\cite{Doi/Edwards:1986,Flory:1949,Fisher:1966,deGennes:1979,%
desCloizeaux/Jannik:1990}, there is no complete agreement about their values on
percolation clusters at
$p_{\mathrm{c}}$~\cite{Nakanishi:1994,Barat/Chakrabarti:1995}. Here, we study
({\it i}\/) the so-called effective coordination number of the cluster, where
contradicting results have been reported using different numerical techniques.
Next, we consider ({\it ii}\/) the enhancement exponent $\gamma$ and ({\it
iii}\/) the exponents $\nu_r$ and $\nu_{\ell}$ characterising the end-to-end
distance of {\it SAW}s in the  $r$- and $\ell$-space metrics. Finally, we
determine ({\it iv}\/) the values of the critical exponents describing the
corresponding structural distribution functions.

We concentrate on {\it SAW}s on percolation clusters at $p_{\mathrm{c}}$ in two
and three dimensions.  In the literature, two distinct methods have been used
for evaluating {\it SAW}s: Exact enumeration ({EE\/}) and Monte Carlo ({MC\/})
simulation. In the {EE\/} technique, {\it all} {\it SAW} configurations on a
given cluster are taken into account, but only relatively short chains can be
evaluated. In a {MC\/} simulation, longer chains can be studied, but inherently
the ensemble of configurations remains incomplete. Here, we use the {EE\/}
technique in combination with an appropriate  finite-size scaling procedure to
determine the relevant exponents. Since `infinitely' long chains can only exist
on the backbone of the cluster, where dangling ends are absent on all length
scales, we study the {\it SAW}s directly on the backbone. This enables us to
generate longer  chains on a given cluster and to average over a larger set of
different cluster configurations.

Specifically, we enumerate all possible {\it SAW} configurations of $N$ steps
for a single backbone and study different moments of the total number of {\it
SAW}s and of their end-to-end distance by averaging over many different
backbone configurations. Our analysis shows that the critical exponents $\nu_r$
and $\nu_{\ell}$ do not depend on the order $q$ of the moments, while the
enhancement exponents and the effective coordination numbers do depend on  $q$,
leading to multifractal behavior. In particular we find that the first moment
of the effective coordination number $\mu_1$ satisfies  $\mu_1 = p_{\mathrm{c}}
\mu$, where $\mu$ is the effective coordination number of the underlying
regular lattice, resolving previous controversies. The mean structural
distribution functions for the end-to-end distance after $N$ steps, both in
Euclidean and topological space, are obtained numerically, supporting the
expected scaling  forms~\cite{Roman/Draeger/Bunde/Havlin/Stauffer:1995,%
Roman/Ordemann/Porto/Bunde/Havlin:1998}.

The paper is organized as follows: In Section~\ref{regular}, we briefly review
the main relevant properties of {\it SAW}s on regular lattices to illustrate
the different numerical procedures employed in this work. In
Section~\ref{percolation}, we present results for the total number and the mean
end-to-end distance of {\it SAW}s on the backbone of the incipient percolation
cluster. The corresponding distribution functions of the end-to-end distance
and their scaling behavior, in Euclidean and topological space, are also
discussed.  Finally, in Section~\ref{conclusion}, we  summarize our main
results.

\section{\textit{SAW}\lowercase{s} on regular lattices reviewed}\label{regular}

In this section we illustrate the different numerical techniques we use in the
following by briefly reviewing the main results for {\it SAW}s on regular
lattices.  The main idea is to show that our finite-size scaling employed in
the later sections enables us to obtain quite accurate estimates for the
critical exponents based on {EE\/} results for relatively short chains. Here,
we consider the case $d=2$, which is particularly suitable since many results
are known exactly.

\subsection{Total number of \textit{SAW} configurations $C_N$}

The total number $C_N$ of {\it SAW} configurations of $N$ steps behaves 
as~\cite{deGennes:1979}
\begin{equation}\label{CN_regular_eq}
C_N = A~\mu^N~N^{\gamma-1}
\end{equation}
where $\mu$ is the effective coordination number of the lattice, $\gamma$ is
the universal enhancement exponent, and $A$ is a constant. To determine $\mu$,
$\gamma$, and $A$, we choose to study the behavior of the quantity
\begin{equation}\label{CNplot_regular_eq}
\frac{\ln C_N}{N} =
\frac{\ln A}{N} + \ln \mu + (\gamma - 1) \, \frac{\ln N}{N}
\end{equation}
as a function of $N$. Fig.~\ref{CN_regular_plot} shows for the square lattice,
that the values for $\mu$ and $\gamma$ obtained by fitting the {EE\/}  data
using Eq.~(\ref{CNplot_regular_eq}) agree well with the accepted values
reported in literature (see Table~\ref{regular_table}).

\subsection{Mean end-to-end distance and structural distribution
function}\label{regular:PrN}

The root mean square end-to-end distance of {\it SAW}s of $N$ steps,
$\overline{r}(N) \equiv [\overline{r^2}(N)]^{1/2} $, averaged over all possible
{\it SAW} configurations behaves as
\begin{equation}\label{r_regular_eq}
\overline{r}(N) \propto N^{\nu_{\rm F}}
\end{equation}
with the universal exponent $\nu_{\rm F} = 3/4$ in $d=2$ as suggested by
Flory~\cite{Flory:1949}.   
In Fig.~\ref{r_regular_plot}, we show  values for
$\overline{r}(N)$ versus $N$ obtained by {EE\/}
technique~\cite{Conway/Guttmann:1996}. The asymptotic value for $\nu_{\rm F}$
(see also Table~\ref{regular_table}) is obtained using successive slopes as
shown in the inset of Fig.~\ref{r_regular_plot} and is in excellent agreement
with the theoretical prediction.

A more detailed information about
the spatial structure of {\it SAW}s is given by the distribution function
$P(r,N)$, where $P(r,N) \, {\rm d} r$ is  the probability that after $N$ steps
the end-to-end distance of a chain is between $r$ and $r + {\rm d} r$. This
quantity obeys the scaling form~\cite{deGennes:1979,desCloizeaux/Jannik:1990}
\begin{equation}\label{PrN_regular_eq}
P(r,N) \propto \frac{1}{r} f(r/N^{\nu_{\rm F}})
\end{equation}
and is normalized according to $\int_0^{\infty} {\rm d} r \, P(r,N) =1$. The
analytic form of the scaling function $f(x)$ is known asymptotically:
\begin{equation}\label{g12_regular_eq}
f(x) \propto
\left\{\begin{array}{ll}
x^{g_1 + d}, & x \ll 1 \\
x^{g_2 + d} \exp \left( -c x^{\delta} \right), & x \gg 1
\end{array}\right. \quad,
\end{equation}
where $g_1 = (\gamma - 1)/\nu_{\rm F}$~\cite{desCloizeaux:1974}, $g_2 = \delta
\, ( d \, [\nu_{\rm F} - 1/2] - [\gamma - 1])$~\cite{McKennzie/Moore:1971} and
$\delta = 1/(1- \nu_{\rm F})$~\cite{Fisher:1966}. Values for these exponents
are summarized in Table~\ref{regular_table}. We have verified these predictions
by enumerating all {\it SAW} configurations for $N = 23$ and $24$ and
calculating the corresponding distributions $P(r,N)$, from which we have
extracted the different exponents (see Fig.~\ref{PrN_regular_plot}). We show
that a more accurate determination of the exponent $g_2$ compared to a simple
fit using  Eq.~(\ref{g12_regular_eq}) can be obtained by employing a specific
numerical procedure described   in Appendix~\ref{wittkop} (see inset of
Fig.~\ref{PrN_regular_plot}). The obtained values are in  agreement with the
theoretical predictions (see Table~\ref{regular_table}).

\section{\textit{SAW}\lowercase{s} on the backbone of the incipient 
percolation cluster}
\label{percolation}

Next, we consider {\it SAW}s on the incipient percolation cluster by generating
all {\it SAW} configurations directly on the backbone of the cluster. We obtain
the backbone of a given cluster grown by the Leath
algorithm~\cite{Leath:1976,Alexandrowics:1980} by randomly choosing {\it one}
of the sites of the last grown cluster shell (e.g.\ site $A$ in
Fig.~\ref{backbone}) and determining the backbone between site $A$ and the seed
of the cluster (site $S$ in Fig.~\ref{backbone}) by the burning procedure
described in~\cite{Herrmann/Hong/Stanley:1984,Porto/Bunde/Havlin/Roman:1997}. 
The {\it SAW}s start at the seed $S$ of the cluster. To avoid finite size
effects, the chemical distance between both endpoints $S$ and $A$ of the
backbone is chosen at least 20 times larger than the chemical length of the
{\it SAW}s. The large ratio between both chemical lengths is needed, since
close to the endpoint $A$, the backbone has a quasi-linear structure, which
would falsify the results for the {\it SAW}s.  The straightforward idea to use
{\it all} sites on the last grown shell as endpoints for the backbone does not
help, but introduces boundary effects in the opposite direction, since in this
case the backbone coincides with the cluster near the end points;
cf.~\cite{Porto/Bunde/Havlin/Roman:1997}.

We analyse the results for {\it SAW}s on the incipient percolation cluster by
applying analogous numerical procedures on the data as described above for {\it
SAW}s on regular lattices. In contrast to the case of regular lattices, on a
percolation cluster two different metrics can be defined, the Euclidean metric
and the topological metric. On average, the chemical distance $\ell$ between
two backbone sites separated by the Euclidean distance $r$ increases with $r$
as~\cite{Pike/Stanley:1981,Havlin/Ben-Avraham:1987}
\begin{equation}\label{dmin_percolation_eq}
\ell \propto r^{d_{\rm min}}
\end{equation}
where $d_{\rm min} = 1.1306 \pm 0.0003$ in $d=2$~\cite{Grassberger:1999} and
$d_{\rm min} = 1.374 \pm 0.004$ in $d=3$~\cite{Grassberger:1993}. Thus
Eq.~(\ref{dmin_percolation_eq}) yields the scaling relation between the two
metrics, which will be used in what follows. Numerically it is found, that data
obtained in $\ell$-space show less fluctuations (cf.\
e.g.~\cite{Roman/Draeger/Bunde/Havlin/Stauffer:1995}). Therefore more accurate
estimates for many characteristic quantities (such as critical exponents) in
$r$-space can be determined by studying the corresponding quantity in
$\ell$-space and transforming it to $r$-space. For example, the  fractal
dimension  of the backbone  in $\ell$-space is $d_{\ell}^{\rm B} = 1.45 \pm
0.01$ in $d=2$ and $d_{\ell}^{\rm B} = 1.36 \pm 0.02$ in $d=3$. Using
Eq.~(\ref{dmin_percolation_eq}), this leads to the values $d_{\rm f}^{\rm B} =
d_{\ell}^{\rm B} d_{\rm min} = 1.64 \pm 0.02$ and $d_{\rm f}^{\rm B} = 1.87 \pm
0.03$ in $r$-space, respectively~\cite{Porto/Bunde/Havlin/Roman:1997}.

\subsection{Total number of \textit{SAW} configurations: Multifractality}

Due to the disordered structure of the clusters, the total number $C_{N,{\rm
B}}$ of {\it SAW} configurations that are generated on a single backbone, with
the seed $S$ of the cluster as starting point, fluctuates strongly among
different backbone configurations. To characterize these fluctuations, we study
the moments ${\left< C^{{q}}_{N,{\rm B}} \right>}$. A similar study on
percolation clusters at criticality has been performed for `ideal' chains,
i.e.\ chains which can intersect themselves. This model leads to a non-trivial
dependence on $q$~\cite{Giacometti/Maritan:1994}.

In generalizing Eq.~(\ref{CN_regular_eq}), we make the ansatz
\begin{equation}\label{CN_percolation_eq}
{\left< C^q_{N,{\rm B}} \right> }^{1/q} =
A_q \, \mu_q^N \, N^{\gamma_q -1} \, ,
\end{equation}
where $\mu_q$ are the generalized effective coordination numbers of the
backbone and $\gamma_q$ the generalized enhancement exponents. Results for
different values of  $q$ are shown in Figs.~\ref{CN_percolation_plot}(a)
and~\ref{CN_percolation_plot}(b) for the square and the simple cubic lattice,
respectively, employing the numerical procedure described in
Section~\ref{regular}A. The values for $\mu_q$ and $\gamma_q$ are displayed in
Fig.~\ref{CN_q_2d_percolation_plot} for $d=2$, clearly revealing a dependence
on $q$, reminiscent of a multifractal behavior.  For large negative values of
$q$, backbone configurations with a small number of {\it SAW} configurations
$C_{N,{\rm B}}$ are singled out in the averaging procedure. We find that $\mu_q
\to 1$ and $\gamma_q \to 1$ for $q \to - \infty$, pointing to rare
configurations of backbones with an almost linear shape. On the contrary, for
large values of $q$ the averaging procedure emphasizes backbone configurations
with a large number of {\it SAW} configurations $C_{N,{\rm B}}$. Since these
backbones are the most compact ones, $\mu_q$ and $\gamma_q$ are strongly
enlarged. Fig.~\ref{CN_q_2d_percolation_plot} seems to suggest that the
structure of the most compact backbones differs distinctively from the
structure of a regular square lattice, as $\lim_{q \to \infty} \mu_q \approx
1.9$, which is well below the value for $\mu$ on the regular square lattice,
and $\lim_{q \to \infty} \gamma_q \approx 1.7$ is well above the value for
$\gamma$ on the regular square lattice.

These results resolve earlier controversies regarding the values for both $\mu$
and $\gamma$ for percolation obtained from {MC\/} simulations and by {EE\/}
techniques. For the square lattice, for example, the values $\mu_{\rm
perc}({\rm EE\/}) = 1.53 \pm 0.05$~\cite{Lam:1990} and $\gamma_{\rm perc}({\rm
EE\/}) = 1.33 \pm 0.02$~\cite{Nakanishi/Lee:1991} have been obtained from exact
enumerations calculations, while from {MC\/} simulations the values $\mu_{\rm
perc}({\rm MC\/}) = 1.459 \pm 0.003$ and $\gamma_{\rm perc}({\rm MC\/}) = 1.31
\pm 0.03$~\cite{Woo/Lee:1991} were determined. We find $\mu_1= 1.565 \pm
0.005$, $\gamma_1= 1.34 \pm 0.05$ and $\mu_0=1.456 \pm 0.005$, $\gamma_0=  1.26
\pm 0.05$, corresponding to the  {EE\/} and {MC\/} results, respectively.  
This can be understood by noting that EE calculations yield by definition the
whole ensemble (the so-called `annealed' average), corresponding to the case
$q=1$, i.e. the normal arithmetic average. In contrast, MC simulations
intrinsicly sample only a small subset of all possible configurations, omitting
rare configurations, yielding `typical' subsets of the ensemble (the so-called
`quenched' average). This quenched average is usually described by a
logarithmic average, i.e. $\left< C_{N,{\rm B}} \right>_{\rm typ} \equiv \exp
\left< \ln C_{N,{\rm B}} \right>$, and is equivalent to the limit $q \to 0$ of
Eq.~(\ref{CN_percolation_eq}), i.e. $\lim_{q \to 0}  \left< C_{N,{\rm B}}^q
\right>^{1/q} = \exp \left< \ln C_{N,{\rm B}}\right>$. Indeed, our results are
in excellent agreement, in both $d=2$ and $d=3$, with the relation
\begin{equation}\label{pc_percolation_eq}
\mu_1 = p_{\rm c} \mu
\end{equation}
where $\mu$ is the effective coordination number of the underlying regular
lattice,  $p_{\rm c} = 0.5927460$ for the square lattice~\cite{Ziff:1994} and
$p_{\rm c} = 0.311605$ for the simple cubic lattice~\cite{Grassberger:1992}.
This relation, which was originally suggested in the form $\mu_{\rm perc} =
p_{\rm c} \mu$~\cite{Woo/Lee:1991}, could not be confirmed earlier on because
of the different values obtained for $\mu_{\rm perc}$.

Because of the possible existence of rare events playing a dominant role in the
average procedure, we have performed a detailed analysis of our numerical data
to confirm that we have considered a sufficiently large set of cluster
configurations (cf.\ Appendix~\ref{generalaverage}).

\subsection{Mean end-to-end distances and structural distribution functions}

Next, we study the scaling behavior of the distribution functions for the
end-to-end distance, $\left< P_{\rm B}(\ell,N) \right>$ and $\left< P_{\rm
B}(r,N) \right>$, averaged over many backbone configurations, where $P_{\rm
B}(\ell,N) \, {\rm d} \ell$  is  the probability that after $N$ steps the
chemical  end-to-end distance of a chain on a single backbone is between $\ell$
and $\ell + {\rm d} \ell$, and $P_{\rm B}(r,N) \, {\rm d} r$ is the analogous
quantity in $r$-space. These distribution functions are expected to obey
scaling forms similar to the one valid on regular lattices,
Eq.~(\ref{PrN_regular_eq}), with the corresponding scaling
exponents~\cite{Roman/Draeger/Bunde/Havlin/Stauffer:1995}. The mean chemical
end-to-end distance $\left< \overline{\ell}(N) \right>$ and the root mean
square Euclidean end-to-end distance $\left< \overline{r}(N) \right> \equiv
\left< \left[ \overline{r^2}(N) \right]^{1/2} \right>$ scale with $N$ as  
\begin{equation}
\left< \overline{\ell}(N) \right> \propto N^{\nu_\ell} 
\end{equation}
and
\begin{equation}
\left< \overline{r}(N) \right> \propto N^{\nu_r},
\end{equation}
respectively. The first average is performed over all {\it SAW} configurations
on a single backbone, the second average is carried out over many backbone
configurations. Following Eq.~(\ref{dmin_percolation_eq}), the exponents
$\nu_\ell$ and $\nu_r$ are related to each other by $\nu_r = \nu_{\ell}/d_{\rm
min}$.  The numerical results for $\nu_\ell$ and $\nu_r$ obtained by the
successive slopes technique discussed in Section~\ref{regular:PrN} for regular
lattices are reported in Table~\ref{percolation_table}. As an example
Fig.~\ref{l_percolation_plot} shows the determination of $\nu_\ell$ in $d=3$.

Accordingly, the scaling variable in chemical space is $\ell/N^{\nu_{\ell}}$
and the mean structural distribution function, averaged over many backbone
configurations, has the form
\begin{equation}\label{PlN_B_eq}
\left< P_{\rm B}(\ell,N) \right> \propto \frac{1}{\ell} f(\ell/N^{\nu_{\ell}})
\quad,
\end{equation}
with the scaling function
\begin{equation}\label{g12l_B_eq}
f_{\ell}(x) \propto
\left\{\begin{array}{ll}
x^{g_1^{\ell} + d_{\ell}^{\rm B}}, & x \ll 1 \\
x^{g_2^{\ell} + d_{\ell}^{\rm B}}
\exp \left( -c_{d,\ell} x^{\delta_{\ell}} \right), & x \gg 1
\end{array}\right. \quad.
\end{equation}
Equivalently, in $r$-space, the scaling variable is $r/N^{\nu_r}$, and one has
\begin{equation}\label{PrN_B_eq}
\left< P_{\rm B}(r,N) \right> \propto \frac{1}{r} f(r/N^{\nu_r})
\end{equation}
with
\begin{equation}\label{g12r_B_eq}
f_r(x) \propto
\left\{\begin{array}{ll}
x^{g_1^r + d_{\rm f}^{\rm B}}, & x \ll 1 \\
x^{g_2^r + d_{\rm f}^{\rm B}}
\exp \left( -c_{d,r} x^{\delta_r} \right), & x \gg 1
\end{array}\right. \quad.
\end{equation}
Both distribution functions are normalized according to $\int_0^{\infty} {\rm
d} \ell \, \left< P_{\rm B}(\ell,N) \right> = 1$ and $\int_0^{\infty} {\rm d} r
\, \left< P_{\rm B}(r,N) \right> = 1$.

The numerical results for the distribution functions in $d=2$ and $d=3$ are
shown in Figs.~\ref{PlrN_2d_percolation_plot}
and~\ref{PlrN_3d_percolation_plot}, respectively, in both $\ell$- and
$r$-space. The values for the exponents $\nu_{\ell}$ and $\nu_r =
\nu_{\ell}/d_{\rm min}$ reported in Table~\ref{percolation_table} are used in
the scaling variables. For the determination of the exponents $g_1^{\ell}$,
$g_2^{\ell}$, $g_1^r$, and $g_2^r$ according to Eqs.~(\ref{g12l_B_eq}) and
(\ref{g12r_B_eq}), we  use the previously reported values of the fractal
dimensions $d_{\ell}^{\rm B}$ and $d_{\rm f}^{\rm
B}$~\cite{Porto/Bunde/Havlin/Roman:1997}. The exponents $g_1^{\ell}$ and
$g_1^r$ can be estimated directly from  the slope of $f_{\ell}$ and $f_{r}$ in
the double logarithmic plots. Since  $g_1^{\ell}$ and $g_1^r$ are related by
$g_1^r = g_1^{\ell} \, d_{\rm
min}$~\cite{Roman/Draeger/Bunde/Havlin/Stauffer:1995}, a more precise estimate
for $g_1^r$ can be derived from the estimate for $g_1^{\ell}$.   The
determination of $g_2^{\ell}$ and $g_2^r$ is more difficult, since both
exponents occur in the non-dominant part and are masked by the exponential.
Therefore it requires the use of the slightly more involved numerical procedure
discussed in Appendix~\ref{wittkop} (see the insets of
Figs.~\ref{PlrN_2d_percolation_plot} and~\ref{PlrN_3d_percolation_plot} for
$d=2$ and $d=3$, respectively). The numerical results we obtain for
$g_1^{\ell}$, $g_2^{\ell}$, $g_1^r$  and $g_2^r$ are reported in
Table~\ref{percolation_table}. Regarding the exponential factors, our results
for the exponents $\delta_{\ell}$ and $\delta_r$ are consistent, within the
present accuracy, with the expressions $\delta_{\ell} = 1/(1-\nu_{\ell})$ and
$\delta_r = 1/(1-\nu_r)$, respectively. 

As discussed in Section~\ref{regular:PrN}, for regular lattices the exponents
$g_1$, $\nu_{\rm F}$ and $\gamma$ are related by the des Cloizeaux relation  
$g_1 = (\gamma-1)/\nu_{\rm F}$. Therefore, it is legitimate to ask if a similar
`generalized des Cloizeaux' relation holds also for {\it SAW}s in percolation.
Since the enhancement exponent $\gamma_q$ depends on $q$, it is necessary to
find out whether the exponents $\nu_{\ell}$ and $g_1^{\ell}$ as well as $\nu_r$
and $g_1^r$ depend on $q$. To this end we generalize the averages $\left<
\overline{\ell}(N) \right>$ and $\left< \overline{r}(N) \right>$ to  $\left<
\overline{\ell}^q(N) \right>^{1/q} \propto N^{\nu_{\ell}^{(q)}}$ and  $\left<
\overline{r}^q(N) \right>^{1/q} \propto N^{\nu_r^{(q)}}$.  Since this is
equivalent of studying the quantities $\left[ \int {\ell}^q  \,  P_{\rm
B}(\ell,N) \, {\rm d} \ell \right]^{1/q}$ and $\left[ \int {r}^q  \,  P_{\rm
B}(r,N) \, {\rm d} r \right]^{1/q}$, respectively, and $ \left< P_{\rm
B}(\ell,N) \right>$ and $ \left< P_{\rm B}(r,N) \right>$ scale with $\ell /
N^{\nu_\ell}$ and $ r / N^{\nu_r}$, also $\left< \overline{\ell}^q(N)
\right>^{1/q} $ and $\left< \overline{r}^q(N) \right>^{1/q} $ must scale with
$\ell / N^{\nu_\ell} $ and $ r / N^{\nu_r} $, respectively.  Therefore
$\nu_\ell^{ (q)} = \nu_\ell $ and $\nu_r^{ (q)} = \nu_r $ for all $q$, which we
confirmed numerically. Furthermore we have numerically verified that the
exponents $g_1^{\ell}$ and $g_1^{r}$ (as well as $g_2^{\ell}$, $g_2^{r}$,
$\delta_{\ell}$ and $\delta_{r}$) are independent of $q$. Regarding the
`generalized des Cloizeaux' relation, our numerical results suggest that in
$d=2$ the relations $g_1^{\ell} =(\gamma_{q=1}-1)/\nu_{\ell} $ and $g_1^r
=(\gamma_{q=1}-1)/\nu_r $ hold, where the, to some extent arbitrary, choice of
$\gamma_{q=1}$ is motivated by the fact that $q=1$ describes the annealed case
of the whole {\it SAW} ensemble. However in $d=3$ these relations are not
satisfied by the present numerical results.  

\section{Concluding remarks}\label{conclusion}

We have studied structural properties of {\it SAW}s on the backbone of the
incipient percolation cluster in both $d=2$ and $d=3$, applying exact
enumeration techniques. Our results suggest that {\it SAW}s display
multifractal behavior,  caused by the underlying multiplicative process 
yielding an infinite hierarchy of generalized coordination numbers $\mu_q$ and
enhancement exponents $\gamma_q$ describing the moments ${\left< C^q_{N,{\rm
B}} \right> }$ of the total number of {\it SAW}s of length $N$. The present
results resolve previous inconsistencies regarding the suggested relation
$\mu_{\rm perc} = p_{\rm c} \mu$, where $p_{\rm c}$ is the percolation
threshold of a specific regular lattice, and $\mu$ and $\mu_{\rm perc}$ are the
corresponding effective coordination numbers of {\it SAW}s for the ordered case
and on the incipient percolation cluster, respectively. We find that this
relation is accurately obeyed on the square and simple cubic lattice by
identifying $\mu_{\rm perc} = \mu_1$.

\section*{Acknowledgements}

We thank A.~Blumen, S.~Edwards and P.~Grassberger for very useful discussions.
Financial support from the German-Israeli Foundation (GIF), the Minerva Center
for Mesoscopics, Fractals, and Neural Networks and the Deutsche
Forschungsgemeinschaft is gratefully acknowledged. M.P. gratefully acknowledges
the Alexander-von-Humboldt foundation (Feodor-Lynen program) for financial
support.

\appendix
\section{Improved procedure for the determination of  $\lowercase{{g_2}}$}
\label{wittkop}

The procedure used for extracting the exponents $g_2$, $g_2^{\ell}$, and
$g_2^{r}$, describing the scaling form of the structural distribution
functions, is an improved version of the procedure by Wittkop {\it et
al.}~\cite{Wittkop/Kreitmeier/Goeritz:1996}
(cf.~\cite{Everaers/Graham/Zuckermann:1995}) and is illustrated here for the
case of regular lattices. The distribution function Eq.~(\ref{PrN_regular_eq})
can be written as
\begin{equation}\label{wittkop_PrN}
P(r,N) = \frac{\Omega B}{r} f(r/N^{\nu_{\mathrm{F}}}) 
\end{equation}
with $\Omega = 2\pi$ in $d=2$ and $\Omega = 4 \pi$ in $d=3$ and the scaling
function $f(x)$ defined as
\begin{equation}\label{wittkop_scaling}
f(x) = \left\{\begin{array}{ll}
x^{g_1+d}, & x < z  \\
x^{g_2+d} \exp \left(-b \, x^{\delta} \right), & x > z
\end{array}\right. \quad,
\end{equation}
where $\delta = 1/(1- \nu_{\rm F})$. The actual value of the crossover $z$
is determined from the numerical results. The constants $B$ and $b$ can be
obtained from the normalization condition
\begin{equation}\label{wittkop_norm1}
\int_0^{\infty} P(r,N) \, {\rm d} r = 1
\end{equation}
and from the second moment
\begin{equation}\label{wittkop_norm2}
\int_0^{\infty}  r^2 \, P(r,N) \, {\rm d} r = \overline{r^2}(N) \cong
N^{2 \nu_{\rm F}}\quad .
\end{equation}
Upon integration of Eqs.~(\ref{wittkop_norm1}) and~(\ref{wittkop_norm2}), one
gets the exact relations
\begin{equation}\label{wittkop_C} 
B = \frac{1}{\Omega}
\left[ \frac{1}{\delta \, b^{(g_2+d)/\delta}}
\Gamma \left( \frac{g_2+d}{\delta}, b z^{\delta} \right) +
\frac{z^{g_1+d}}{g_1+d} \right]^{-1} \quad ,
\end{equation} 
where $\Gamma(u,z)$ is the incomplete Gamma function, and
\begin{equation}\label{wittkop_c} 
\Omega B \left\{ \frac{1}{\delta \, b^{(g_2+d+2)/\delta}}
\Gamma \left( \frac{g_2+d+2}{\delta}, b z^{\delta} \right) \; +
\frac{z^{g_1+d+2}}{g_1+d+2} \right\} = 1 \qquad .
\end{equation}

Thus, by plotting the distribution function in the case $x > z$ as $y \equiv
b^{(g_2 +d)/\delta} (\Omega B)^{-1} \, r \, P(r,N) \exp \left[ \left(
b^{1/\delta} r/N^{\nu_{\mathrm{F}}} \right)^{\delta} \right] $ versus $
b^{1/\delta} \, r/N^{\nu_{\mathrm{F}}}$ in a double logarithmic plot, the
exponent $g_2$ can be read off from the relation $y \sim x^{g_2 +d}$ and
adjusted until the above relations Eqs.~(\ref{wittkop_C}) and~(\ref{wittkop_c})
are satisfied. This method yields much more accurate results than by directly
fitting the distribution function itself. The accuracy of the result can be
assessed by plotting $y \equiv -\ln \left[ b^{(g_2 +d)/\delta} (\Omega B)^{-1}
r \, P(r,N) \left( b^{1/\delta} \, r/N^{\nu_{\mathrm{F}}} \right)^{-(g_2+d)}
\right]= b \, \left(  r/N^{\nu_{\mathrm{F}}} \right)^{\delta}$ versus $
b^{1/\delta} \, r/N^{\nu_{\mathrm{F}}}$ in a double logarithmic plot, from
which the exponent $\delta$ can be determined and compared with the expected
value $\delta = 1/(1 - \nu_{\mathrm{F}})$. The procedure can be extended
straightforwardly to study the distribution functions $\left< P_{\rm B}(\ell,N)
\right>$ and $\left< P_{\rm B}(r,N) \right>$ of {\it SAW}s on the backbone of
critical percolation clusters.

\section{Generalized averaging procedure}\label{generalaverage}

To obtain an estimate of whether the ensemble ${\cal B}$  of backbone
configurations considered is sufficiently large in order to get convergent
results, we analyse the data by a generalized averaging procedure as follows:
The total ensemble ${\cal B}$ containing $n_{\rm tot}$ backbone configurations
is divided into subsets ${\cal B}_i$ containing $n_{\rm eff}$ configurations
each. The generalized average is then defined as
\begin{equation}
{\left< {C_{N,{\rm B}}} \right>}_{n_{\rm eff}}^{(q)}=
\left( \frac{1}{n_{\rm eff}}
\sum_{i = 1}^{n_{\rm eff}} ({C_{N,{\rm B}}})_i^q \right)^{1/q} \qquad .
\end{equation}
The obtained results ${\left< {C_{N,{\rm B}}} \right>}_{n_{\rm eff}}^{(q)}$
depend sensitively on the different backbone configurations and display strong
fluctuations, indicating that the system is not self-averaging. In order to
smooth out these fluctuations, a second average is performed. This second step
is a logarithmic average over the $n_{\rm tot} / n_{\rm eff}$
subsets~\cite{Bunde/Draeger:1995}
\begin{equation}\label{logaverage_eq}
C_{N,{\rm B}}(q,n_{\rm eff}) = \exp \left<{ \ln
 \left< C_{N,{\rm B}} \right>_{n_{\rm eff}}^{(q)}} \right> \\
 {} = A_{q,n_{\rm eff}} 
 {\mu}_{q,n_{\rm eff}}^N \; N^{\gamma_{q,n_{\rm eff}}-1} \qquad .
\end{equation}

In Eq.~(\ref{logaverage_eq}), the limiting case $n_{\rm eff} = 1$ corresponds
to the limit $q \to 0$, while the usual average (cf.\
Eq.~(\ref{CN_percolation_eq})) is recovered when $n_{\rm eff}= n_{\rm tot}$.
The results for the coordination numbers $\mu_{q,n_{\rm eff}}$ and enhancement
exponents $\gamma_{q,n_{\rm eff}}$ are shown in Figs.~\ref{appendix}(a) (for
$q=1$) and~\ref{appendix}(b) (for $q=2$). A dependence of these two values on
$n_{\rm eff}$ indicates that the given ensemble is too small to obtain the
asymptotic values. If, on the contrary, the ensemble of backbone configurations
is sufficiently large, then $\mu_{q,n_{\rm eff}}$ and $\gamma_{q,n_{\rm eff}}$
no longer depend on $n_{\rm eff}$. For $q=1$, this seems to be the case when
$n_{\rm eff} \gtrsim 10^3$, and for $q=2$ when $n_{\rm eff} \gtrsim 10^4$.


\newpage

\def\citeforcite#1{\raisebox{1ex}{{\scriptsize #1}}\,\ignorespaces}
\begin{table}[t]
\begin{tabular}{c||r@{}l|r@{}l}
 & \multicolumn{2}{c}{literature} & \multicolumn{2}{c}{present results} \\
\hline & & & & \\[-8.0mm] \hline
$\gamma$ & $43/32$ & \citeforcite{a} & $1.30 \pm 0.05 $ & \\
\hline
$\mu$ & $2.6385 \pm 0.0001$ & \citeforcite{b} & $2.641 \pm 0.005$ & \\
\hline
$\nu_{\rm F}$ & $3/4$ & \citeforcite{c} & $0.745 \pm 0.005$ & \\
\hline
$g_1$ & $11/24$ & \citeforcite{d} & $0.4 \pm 0.1$ & \\
\hline
$g_2$ & $5/8$ & \citeforcite{e} & $0.61 \pm 0.05$ & \\
\hline
$\delta$ & $4$ & \citeforcite{f} & $4.5 \pm 0.5$ & \\
\end{tabular}
\caption{Structural parameters for {\it SAW}s on regular lattices in $d=2$.
Results of the present simulations obtained on the square lattice, compared
with the accepted values from the literature:
\protect\citeforcite{a}Ref.~\protect\cite{Niehnhuis:1982},
\protect\citeforcite{b}Ref.~\protect\cite{Guttmann/Wang:1991,%
Masand/Wilensky/Massar/Redner:1992},
\protect\citeforcite{c}Ref.~\protect\cite{Flory:1949},
\protect\citeforcite{d}Ref.~\protect\cite{desCloizeaux:1974},
\protect\citeforcite{e}Ref.~\protect\cite{McKennzie/Moore:1971}, and
\protect\citeforcite{f}Ref.~\protect\cite{Fisher:1966}.}
\label{regular_table}
\end{table}

\begin{table}[t]
\begin{tabular}{c||r|r}
 & $d=2$ & $d=3$ \\
\hline & & \\[-8.0mm] \hline
$\gamma_1$ & $1.34 \pm 0.05$ & $1.29 \pm 0.05 $ \\
\hline
$\gamma_0$ & $1.26 \pm 0.05$ & $1.19\pm 0.05 $ \\
\hline
$\mu_1$ & $1.565 \pm 0.005$ & $1.462 \pm 0.005$ \\
\hline
$\mu_0$ & $1.456 \pm 0.005$ & $1.317 \pm 0.005$ \\
\hline
$\nu_{\ell}$ & $0.89 \pm 0.01$ & $0.910 \pm 0.005$ \\
\hline
$\nu_{r}$ (directly from data) & $0.778 \pm 0.015$ & $0.66 \pm 0.01$ \\
\hline
$\nu_{r} = \nu_{\ell}/d_{\rm min}$ & $0.787\pm 0.010$ & $0.662 \pm 0.006$ \\
\hline
$g_1^{\ell}$ & $0.45 \pm 0.10$ & $0.66 \pm 0.15$ \\
\hline
$g_1^r = g_1^{\ell} \, d_{\rm min}$ & $0.51 \pm 0.11$ & $0.91 \pm 0.20$ \\
\hline
$g_2^{\ell}$ & $1.6 \pm 0.16$ & $1.95 \pm 0.17$ \\
\hline
$g_2^r$ & $1.26 \pm 0.18$ & $2.96 \pm 0.18$ \\
\hline
$\delta_{\ell}$ & $9.5 \pm 0.5$ & $12 \pm 0.5$ \\
\hline
$\delta_{r}$ & $4.85 \pm 0.20$ & $3.1 \pm 0.2$ \\
\end{tabular}
\caption{Structural parameters for {\it SAW}s on the backbone of percolation
clusters at criticality in $d=2$ and $d=3$, on the square and simple cubic
lattice, respectively.  The values for $\nu_r$ obtained directly from the
numerical data are in agreement with the more precise values obtained from the
relation  $\nu_{r} = \nu_{\ell}/d_{\rm min}$. The values for $g_1^r =
g_1^{\ell} \; d_{\rm min}$ are also in  good agreement with the corresponding
values obtained directly from the data.  The numerical values for the exponents
$g_2^{\ell}$ and $g_2^r$  have been determined using the  procedure described
in Appendix~\ref{wittkop}. Note that there is no simple relation between 
$g_2^{\ell}$ and $g_2^r$, i.e.\ $g_2^r \neq g_2^{\ell} \; d_{\rm min}$.  The
values of $\delta_{\ell}$ and $\delta_r$ are consistent, within the present
accuracy,  with the expressions $\delta_{\ell} = 1/(1-\nu_{\ell})$ and
$\delta_r = 1/(1-\nu_r)$.}
\label{percolation_table}
\end{table}

\newpage\hbadness 3000

\begin{figure}
\caption{The total number $C_N$ of {\it SAW} configurations on the square
lattice plotted as $(\ln C_N) /N$ versus $N$, from the presently available
exact enumeration results for $C_N$, $N \le 51$
\protect\cite{Conway/Guttmann:1996}. The continuous line corresponds to a
numerical fit obtained in the range $10 \le N \le 51$ using
Eq.~(\ref{CNplot_regular_eq}), with $\mu = 2.641$, $\gamma = 1.3$ and $A =
1.35$.}
\label{CN_regular_plot}
\end{figure}

\begin{figure}
\caption{The mean end-to-end distance $\overline{r}(N)$ versus $N$ for {\it
SAW}s on the square lattice. The continuous line is drawn as a guide and its
slope has the theoretical value $\nu_{\rm F} = 3/4$. In the inset, the
successive slopes $\nu_{\rm F} = {\rm d}\ln \overline{r}(N)/{\rm d}\ln N$ are
plotted versus $1/N$. A linear extrapolation of the points to the limit $1/N
\rightarrow 0$ yields our estimate $\nu_{\rm F} = 0.745 \pm 0.005$, consistent
with the value $3/4$.}
\label{r_regular_plot}
\end{figure}

\begin{figure}
\caption{The structural distribution function of {\it SAW}s, $r \, P(r,N)$
versus $r/N^{\nu_{\rm F}}$, with $\nu_{\rm F} = 3/4$, for $N=23$~(diamonds) and
$N=24$~(circles) on the square lattice. The dashed line in the range
$r/N^{\nu_{\rm F}} < 1$ has a slope $g_1 + d = 2.4$, and the one for
$r/N^{\nu_{\rm F}} > 1$ is a fit with Eq.~(\ref{g12_regular_eq}), for $x \gg
1$, yielding $g_2 + d = 2.9 \pm 0.4$, $\delta = 4.5 \pm 0.8$, and $c = 0.7 \pm
0.1$. In the inset, we show the function $r \tilde{P}(r,N) \equiv  b^{(g_2
+d)/\delta} (\Omega B)^{-1} \, r \, P(r,N) \exp \left[ \left( b^{1/\delta}
r/N^{\nu_{\mathrm{F}}} \right)^{\delta} \right] = b \, \left( 
r/N^{\nu_{\mathrm{F}}} \right)^{\delta} $ versus $ b^{1/\delta} r/N^{\nu_{\rm
F}}$, following the procedure described in Appendix~\ref{wittkop}, allowing a
more precise determination of $g_2$. For our estimate of the crossover value
$z=0.4$ the continuous line has a slope $g_2 + d = 2.61 \pm 0.05$, in agreement
with the theoretical value (see Table~\ref{regular_table}).}
\label{PrN_regular_plot}
\end{figure}

\begin{figure}
\caption{A percolation cluster on the square lattice (full squares) and its
corresponding backbone between the seed $S$ and a  site $A$ randomly chosen on
the last grown shell.}
\label{backbone}
\end{figure}

\begin{figure}
\caption{Generalized moments $\left< C_{N,{\rm B}}^q \right>$ of the
total number $C_{N,{\rm B}}$ of {\it SAW} configurations on the backbone of
critical percolation clusters, plotted as $(1/N) \ln\left[\left< C_{N,{\rm
B}}^q \right>^{1/q}\right]$ versus $N$. {\bf (a)}~On the square lattice, for $q
= 2$, $1$, $0.5$, $0$, $-0.5$, $-1$ and $-2$ (from top to bottom); {\bf (b)}~on
the simple cubic lattice, for $q = 1$ (top) and $q=0$ (bottom). Averages over
$10^5$ backbone configurations each are performed. The continuous lines are
the best fits based on Eq.~(\ref{CN_percolation_eq}), yielding the values for
$\mu_q$ and $\gamma_q$ for $q=0$ and $1$ given in
Table~\protect\ref{percolation_table}. Some representative values for
$\gamma_q$, in addition to those reported in
Table~\protect\ref{percolation_table}, are: $\gamma_{-2} = 1.15\pm 0.05$,
$\gamma_{-1} = 1.23\pm 0.05$, and $\gamma_{2} = 1.36\pm 0.05$ in $d=2$. Values
of $A_q$ are found to fluctuate in the range of $1.0$ to $1.3$ in both $d=2$ 
and $d=3$.}
\label{CN_percolation_plot}
\end{figure}

\begin{figure}
\caption{The effective coordination numbers $\mu_q$ and enhancement exponents
$\gamma_q$ versus $q$ for $-10 \le q \le 10$ in $d=2$ obtained from
Fig.~\protect\ref{CN_percolation_plot}(a).  Except for $\gamma_q$ for $q \ge 2$
the errorbars are smaller than the symbolsizes. The values for $\mu$ and
$\gamma$ on regular square lattice are marked by arrows, clearly showing that 
$\lim_{q \to \infty} \gamma_q $ is larger than $\gamma$ on regular square
lattice.  The inset shows $\mu_q$ versus $q$ for $-2 \le q \le 2$ in $d=2$, in
good agreement with the theoretical result $\mu_q = \mu_0 (1 + q \sigma_0^2/2)$
(continuous line), expected for $\vert q \vert \to 0$
\protect\cite{Roman/Ordemann/Porto/Bunde/Havlin:1998}, with $\mu_0 = 1.456$ and
$\sigma_0 = 0.45$.}
\label{CN_q_2d_percolation_plot}
\end{figure}

\begin{figure}
\caption{The mean topological end-to-end distance $\left< \overline{\ell}(N)
\right>$ versus $N$ for {\it SAW}s  on the backbone of critical percolation
clusters in $d=3$  averaged over $5 \cdot 10^4$ backbone configurations. In the
inset, the successive slopes $\nu_{\ell} = {\rm d}\ln \left< \overline{\ell}(N)
\right> /{\rm d}\ln N$ are plotted versus $1/N$. A linear extrapolation of the
points to the limit $1/N \rightarrow 0$ yields our estimate $\nu_{\ell} = 0.910
\pm 0.005$. }
\label{l_percolation_plot}
\end{figure}

\begin{figure}
\caption{Scaling plots of the distribution functions on the backbone in $d=2$,
for $N=39$ and $40$, averaged over $5 \cdot 10^3$ configurations. {\bf
(a)}~$\ell \, \left< P_{\rm B}(\ell,N) \right>$ versus $\ell/N^{\nu_{\ell}}$:
The dashed line has the slope $1.90$ and corresponds to the ansatz
Eq.~(\ref{g12l_B_eq}), for $x \ll 1$; the continuous line is a fit with the
ansatz Eq.~(\ref{g12l_B_eq}), for $x \gg 1$, yielding $g_2^{\ell} = 1.4 \pm
0.4$, $\delta_{\ell} = 9.5 \pm 0.5$, and $c_{2,\ell} = 0.09 \pm 0.01$. The
inset shows $\ell \left< \tilde{P}_{\rm B}(\ell,N) \right> \equiv
b_{\ell}^{(g_2^{\ell} +d_{\ell}^{\rm B})/\delta_{\ell}} (\Omega B_{\ell})^{-1}
\, \ell \, \left< P_{\rm B}(\ell,N) \right> \exp \left[ \left(
b_{\ell}^{1/\delta_{\ell}} {\ell}/N^{\nu_{\ell}} \right)^{\delta_{\ell}}
\right] = b_{\ell} \, \left(  {\ell}/N^{\nu_{\ell}} \right)^{\delta_{\ell}} $
versus $ b_{\ell}^{1/\delta_{\ell}} \, {\ell}/N^{\nu_{\ell}}$, with our
estimate of the crossover value $z_{\ell}=0.21$,  according to the procedure
described in Appendix~\ref{wittkop}, yielding the more precise estimate
$g_2^{\ell} + d_{\ell}^{\rm B} = 3.05 \pm 0.15$ (continuous line). 
{\bf (b)}~$r  \, \left< P_{\rm B}(r,N) \right>$ versus $r/N^{\nu_{r}}$: The
dashed line has the slope $2.15$ and corresponds to the ansatz
Eq.~(\ref{g12r_B_eq}), for $x \ll 1$; the continuous line is a fit with the
ansatz Eq.~(\ref{g12r_B_eq}), for $x \gg 1$, yielding $g_2^{r} = 1.46 \pm 0.4$,
$\delta_{r} = 4.9 \pm 0.3$, and $c_{2,r} = 0.79 \pm 0.10$. The inset shows $r
\left< \tilde{P}_{\rm B}(r,N) \right> \equiv b_{r}^{(g_2^r +d_r^{\rm
B})/\delta_r} (\Omega B_r)^{-1} \, r \, \left< P_{\rm B}(r,N) \right> \exp
\left[ \left( b_{r}^{1/\delta_{r}} {r}/N^{\nu_{r}} \right)^{\delta_{r}} \right]
= b_r \, \left(  r/N^{\nu_{r}} \right)^{\delta_{r}}$ versus $
b_{r}^{1/\delta_{r}} \, {r}/N^{\nu_{r}}$ with our estimate of the crossover
value $z_r=0.25$, according to the procedure described in
Appendix~\ref{wittkop}, yielding the more precise estimate $g_2^{r} +
d_{r}^{\rm B} = 2.9 \pm 0.15$ (continuous line).}
\label{PlrN_2d_percolation_plot}
\end{figure}

\begin{figure}
\caption{Scaling plots of the distribution functions on the backbone in $d=3$,
for $N=39$ and $40$, averaged over $5 \cdot 10^3$ configurations. {\bf
(a)}~$\ell \,  \left< P_{\rm B}(\ell,N) \right>$ versus $\ell/N^{\nu_{\ell}}$:
The dashed line has the slope $2.02$ and corresponds to the ansatz
Eq.~(\ref{g12l_B_eq}), for $x \ll 1$; the continuous line is a fit with the
ansatz Eq.~(\ref{g12l_B_eq}), for $x \gg 1$, yielding $g_2^{\ell} = 1.3 \pm
0.6$, $\delta_{\ell} = 12.0 \pm 0.5$, and $c_{3,\ell} = 0.06 \pm 0.01$. The
inset shows $\ell \left< \tilde{P}_{\rm B}(\ell,N) \right> \equiv
b_{\ell}^{(g_2^{\ell} +d_{\ell}^{\rm B})/\delta_{\ell}} (\Omega B_{\ell})^{-1}
\, \ell \, \left< P_{\rm B}(\ell,N) \right> \exp \left[ \left(
b_{\ell}^{1/\delta_{\ell}} {\ell}/N^{\nu_{\ell}} \right)^{\delta_{\ell}}
\right] = b_{\ell} \, \left(  {\ell}/N^{\nu_{\ell}} \right)^{\delta_{\ell}}$
versus $ b_{\ell}^{1/\delta_{\ell}} \, {\ell}/N^{\nu_{\ell}}$, with our
estimate of  the crossover value $z_{\ell}=0.4$, according to the procedure
described in Appendix~\ref{wittkop}, yielding the more precise estimate
$g_2^{\ell} + d_{\ell}^{\rm B} = 3.31 \pm 0.15$ (continuous line). 
{\bf (b)}~$r  \, \left< P_{\rm B}(r,N) \right>$ versus $r/N^{\nu_{r}}$: The
dashed line has the slope $2.78$ and corresponds to the ansatz
Eq.~(\ref{g12r_B_eq}), $x \ll 1$; the continuous line is a fit with the ansatz
Eq.~(\ref{g12r_B_eq}), $x \gg 1$, yielding $g_2^{r} = 2.3 \pm 0.6$, $\delta_{r}
= 3.5 \pm 0.5$, and $c_{3,r} = 0.88 \pm 0.10$. The inset shows $r \left<
\tilde{P}_{\rm B}(r,N) \right> \equiv b_{r}^{(g_2^r +d_r^{\rm B})/\delta_r}
(\Omega B_r)^{-1} \, r \, \left< P_{\rm B}(r,N) \right> \exp \left[ \left(
b_{r}^{1/\delta_{r}} {r}/N^{\nu_{r}} \right)^{\delta_{r}} \right] = b_r \,
\left(  r/N^{\nu_{\ell}} \right)^{\delta_{\ell}}$ versus $ b_{r}^{1/\delta_{r}}
\, {r}/N^{\nu_{r}}$ with our estimate of the crossover value $z_r=0.5$,
yielding the more precise estimate $g_2^{r} + d_{r}^{\rm B} = 4.83 \pm 0.15$
(continuous line).}
\label{PlrN_3d_percolation_plot}
\end{figure}

\begin{figure}
\caption{The effective coordination numbers $\mu_{q, n_{\rm eff}}$~(circles)
and the enhancement exponents $\gamma_{q, n_{\rm eff}}$~(squares) of {\it SAW}s
on the backbone in $d=2$ for: {\bf (a)}~$q=1$ and {\bf (b)}~$q=2$ versus the
effective ensemble size $n_{\rm eff}$. The values are obtained by a
least-square-fit of $(\ln [{C_{N, {\rm B}} (q, n_{\rm eff})} ])/N = (\ln A_{q,
n_{\rm eff}})/N + \ln \mu_{q, n_{\rm eff}} + [(\gamma_{q, n_{\rm eff}} - 1) \,
\ln N]/N$ versus $N$, shown as $\mu_{q, n_{\rm eff}}$ and $\gamma_{q, n_{\rm
eff}}-1$ versus $n_{\rm eff}$.} \label{appendix}
\end{figure}

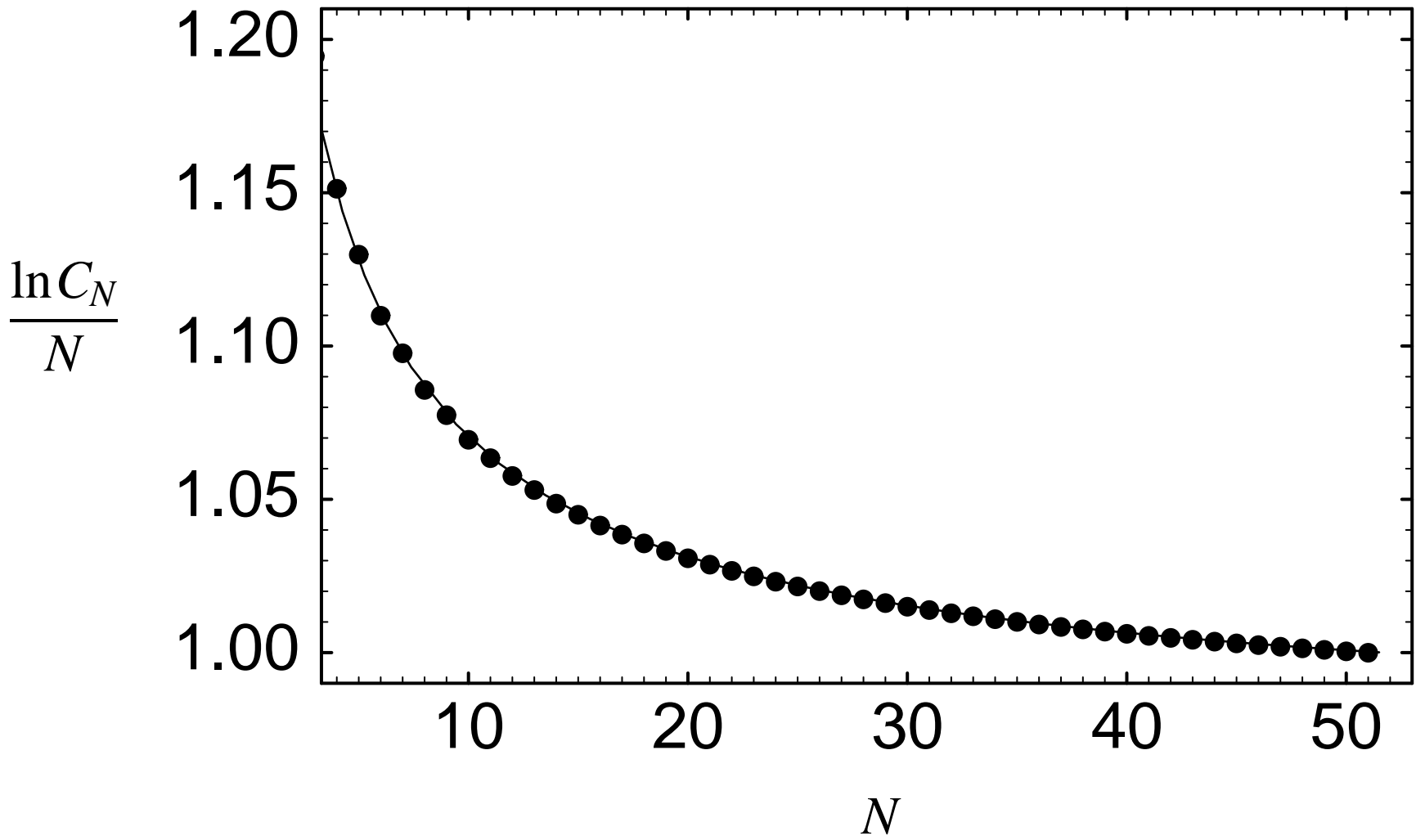
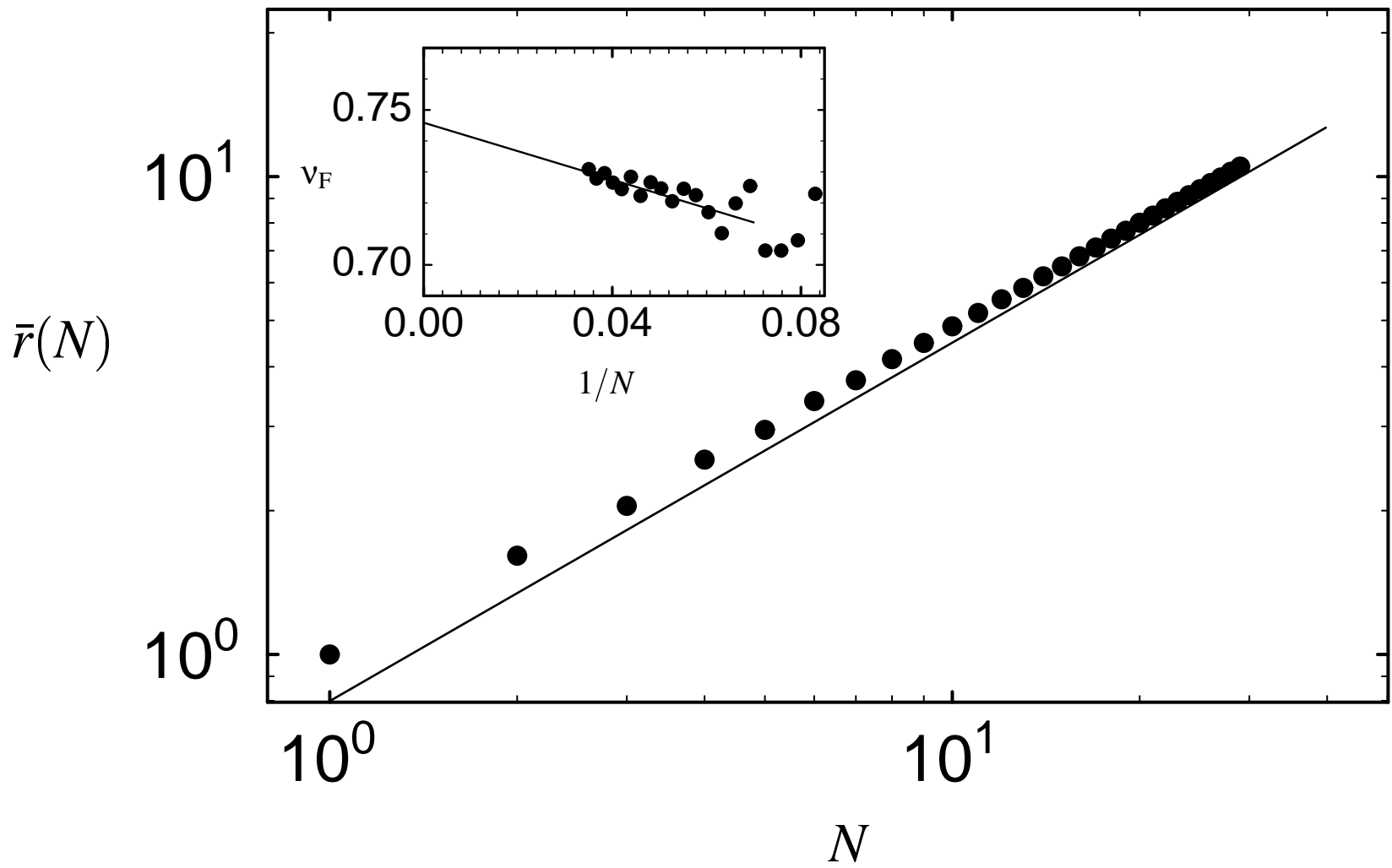
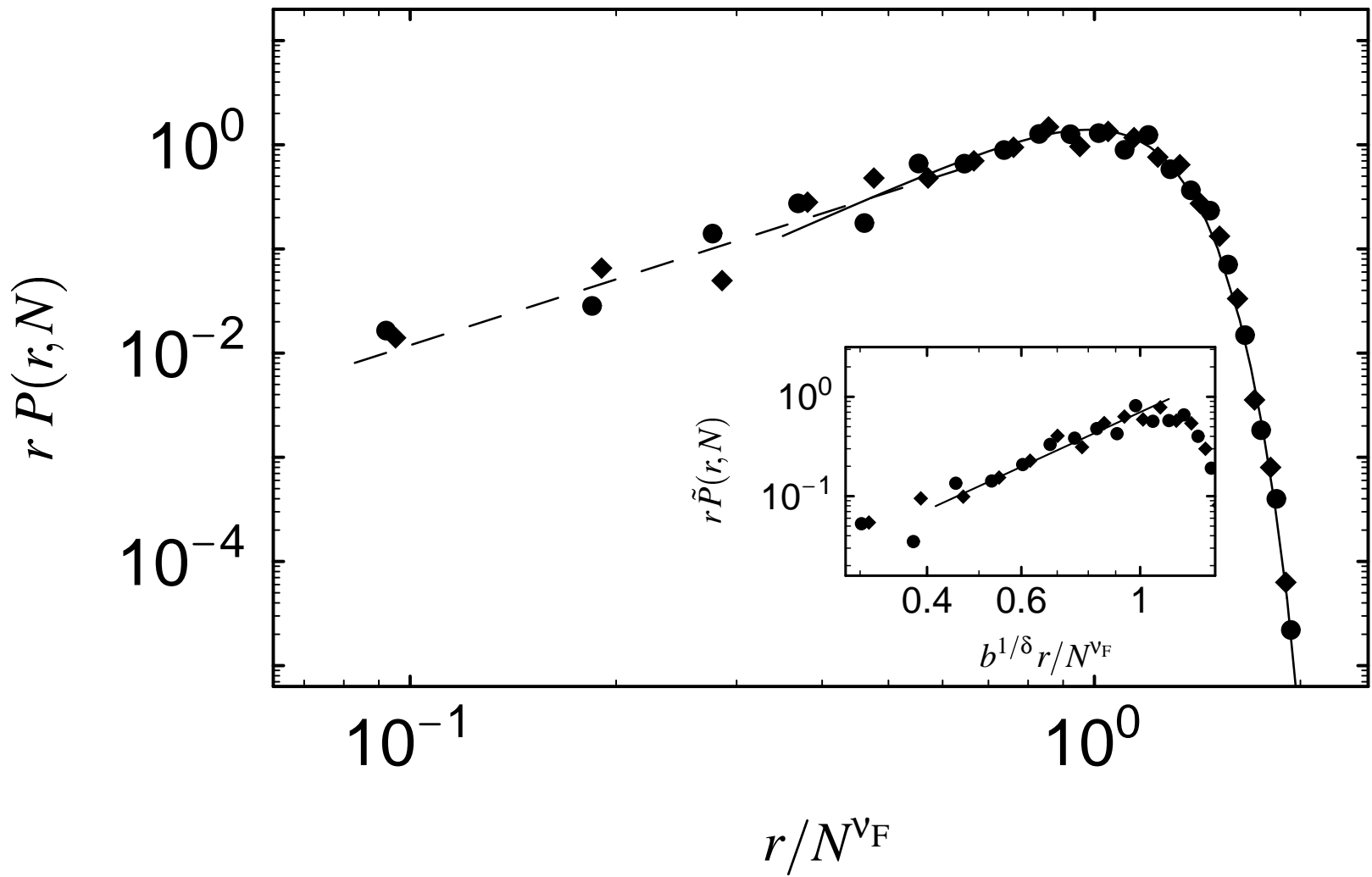
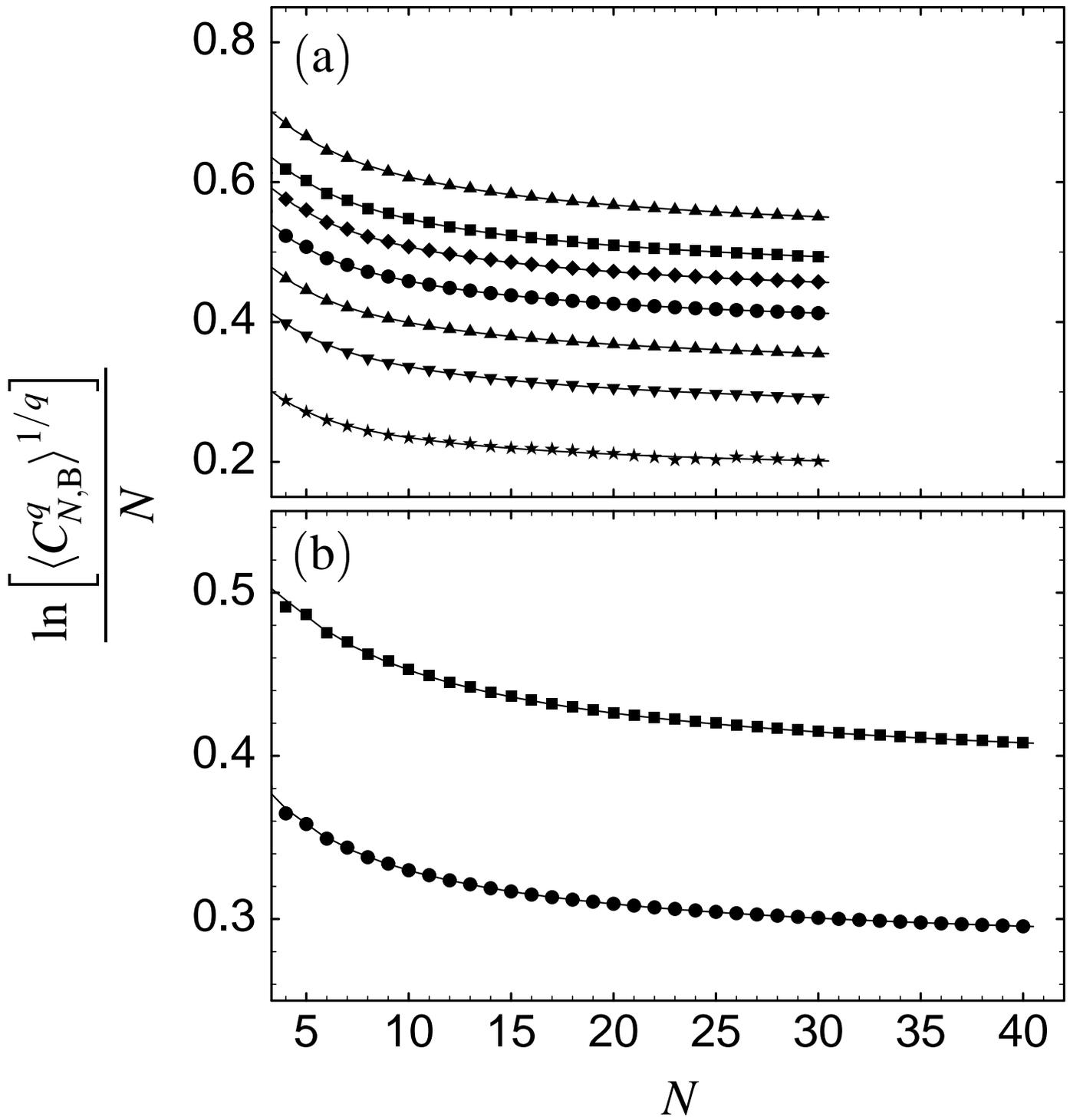
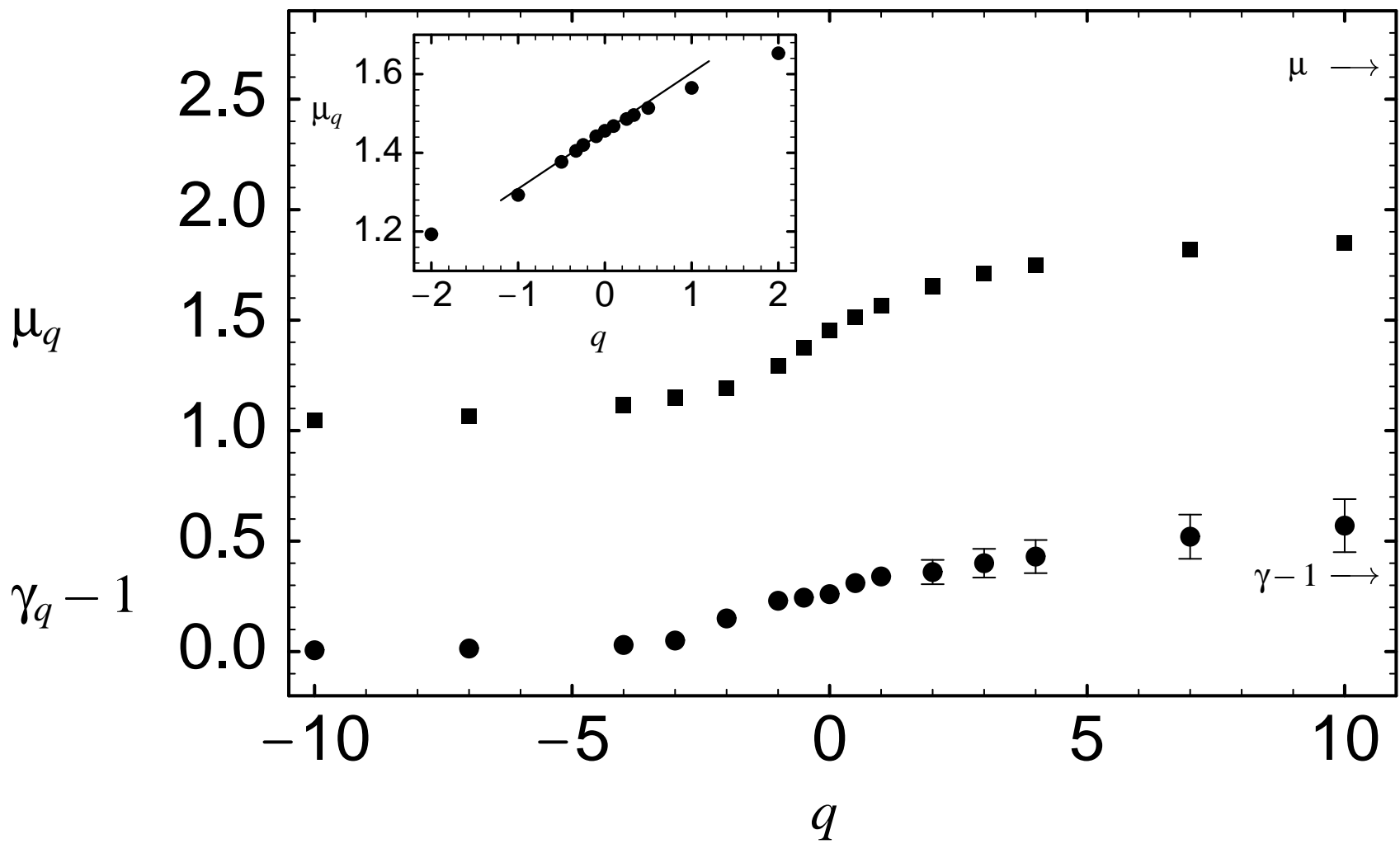
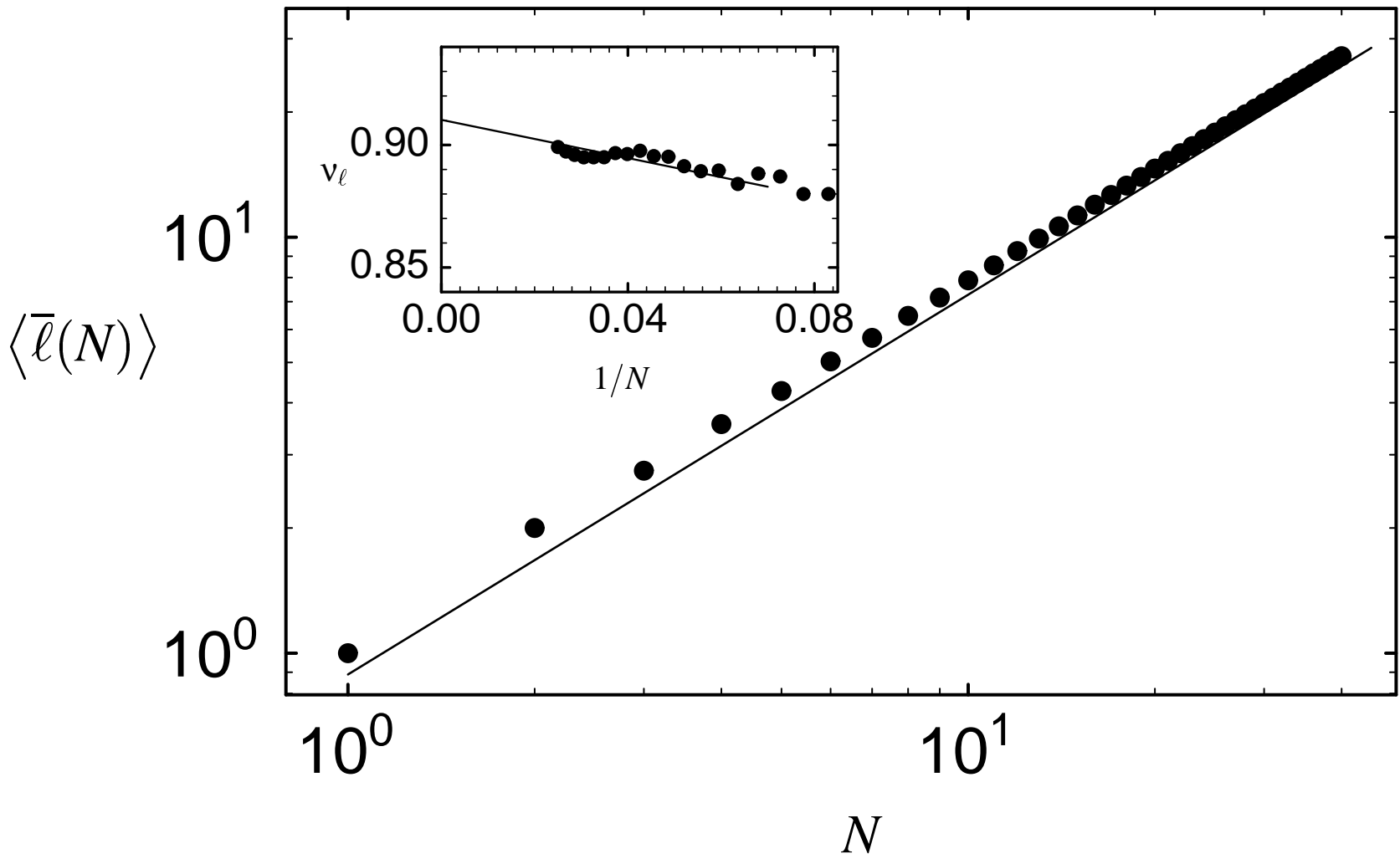
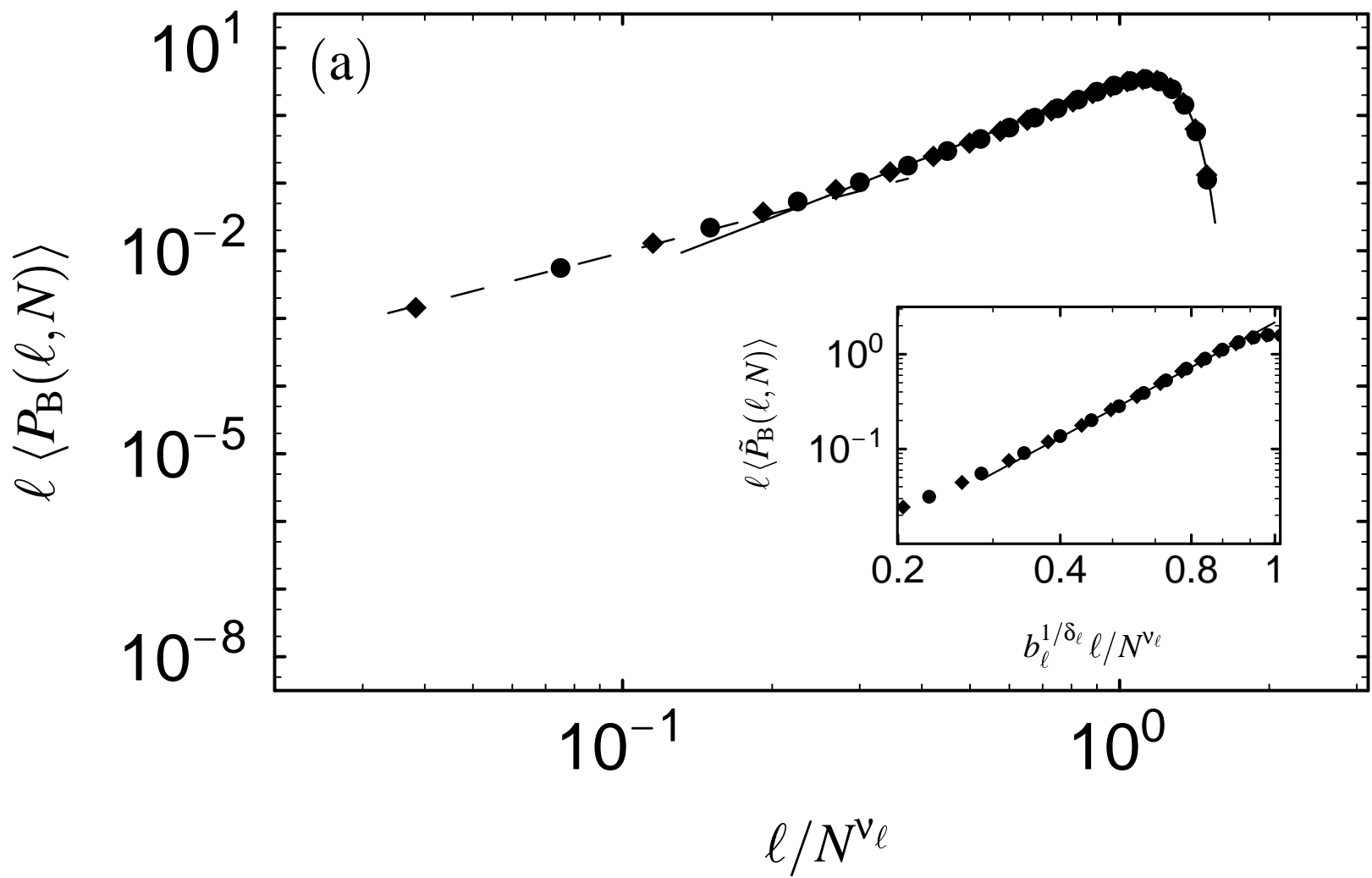
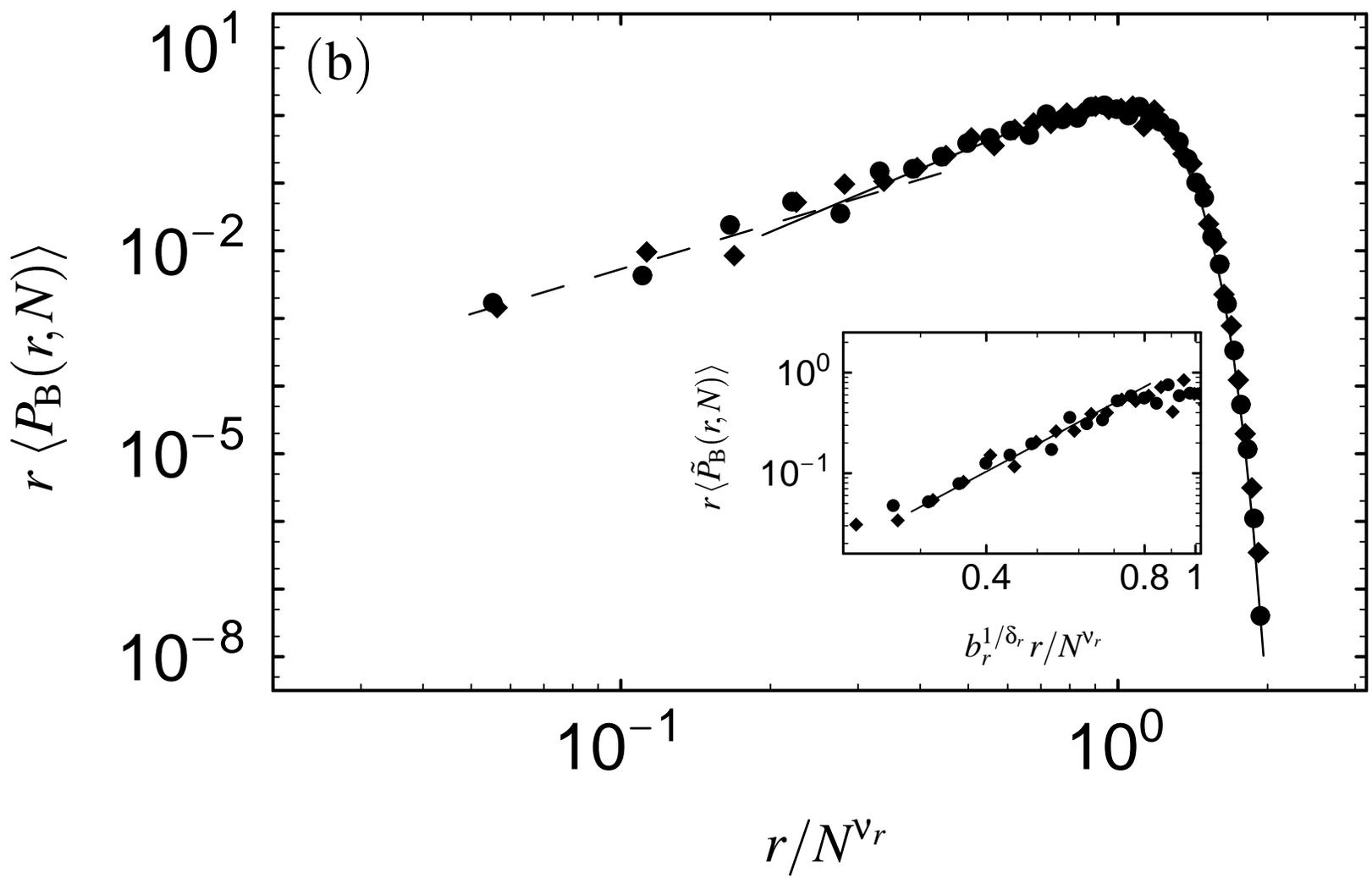
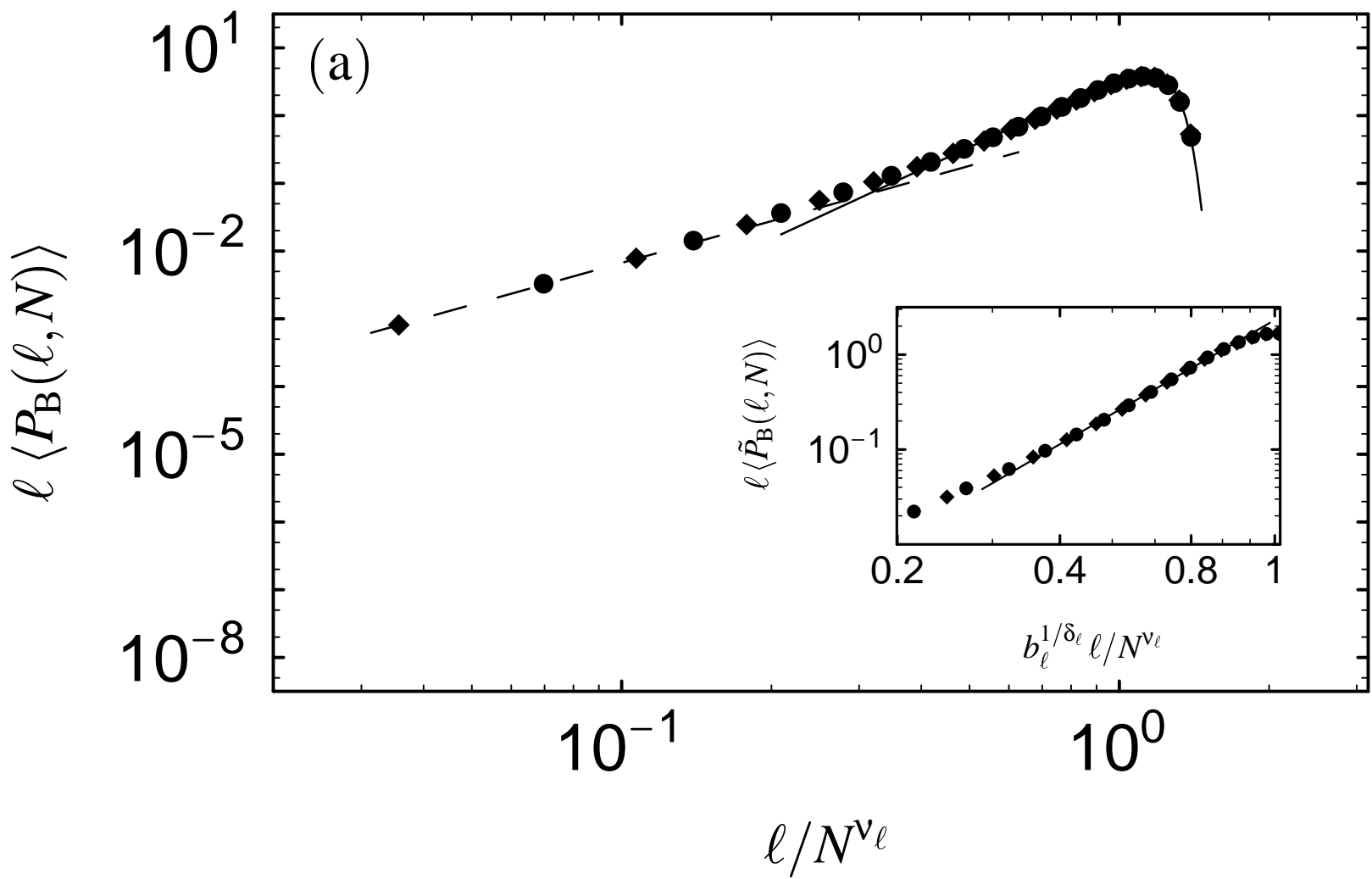
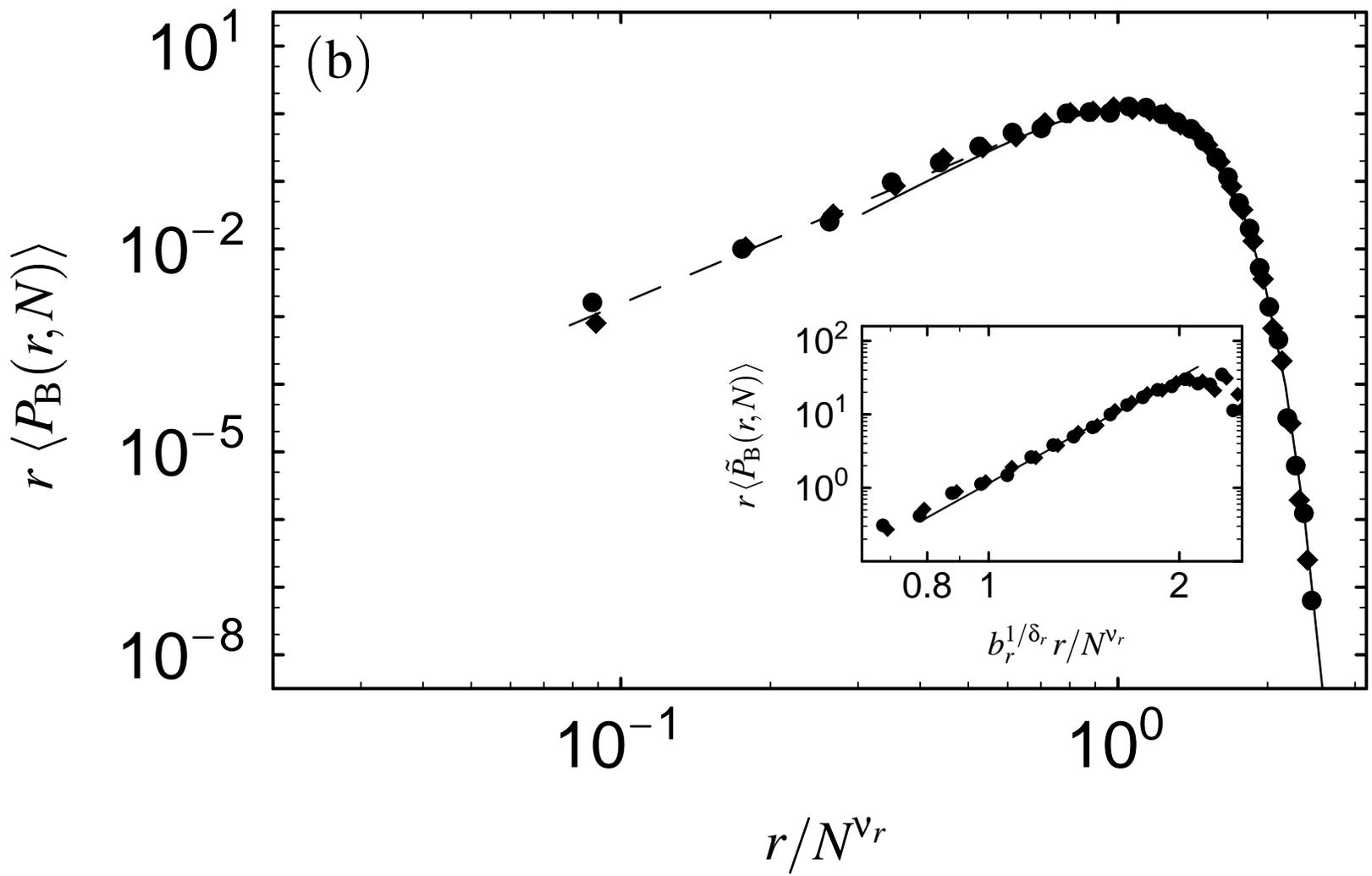
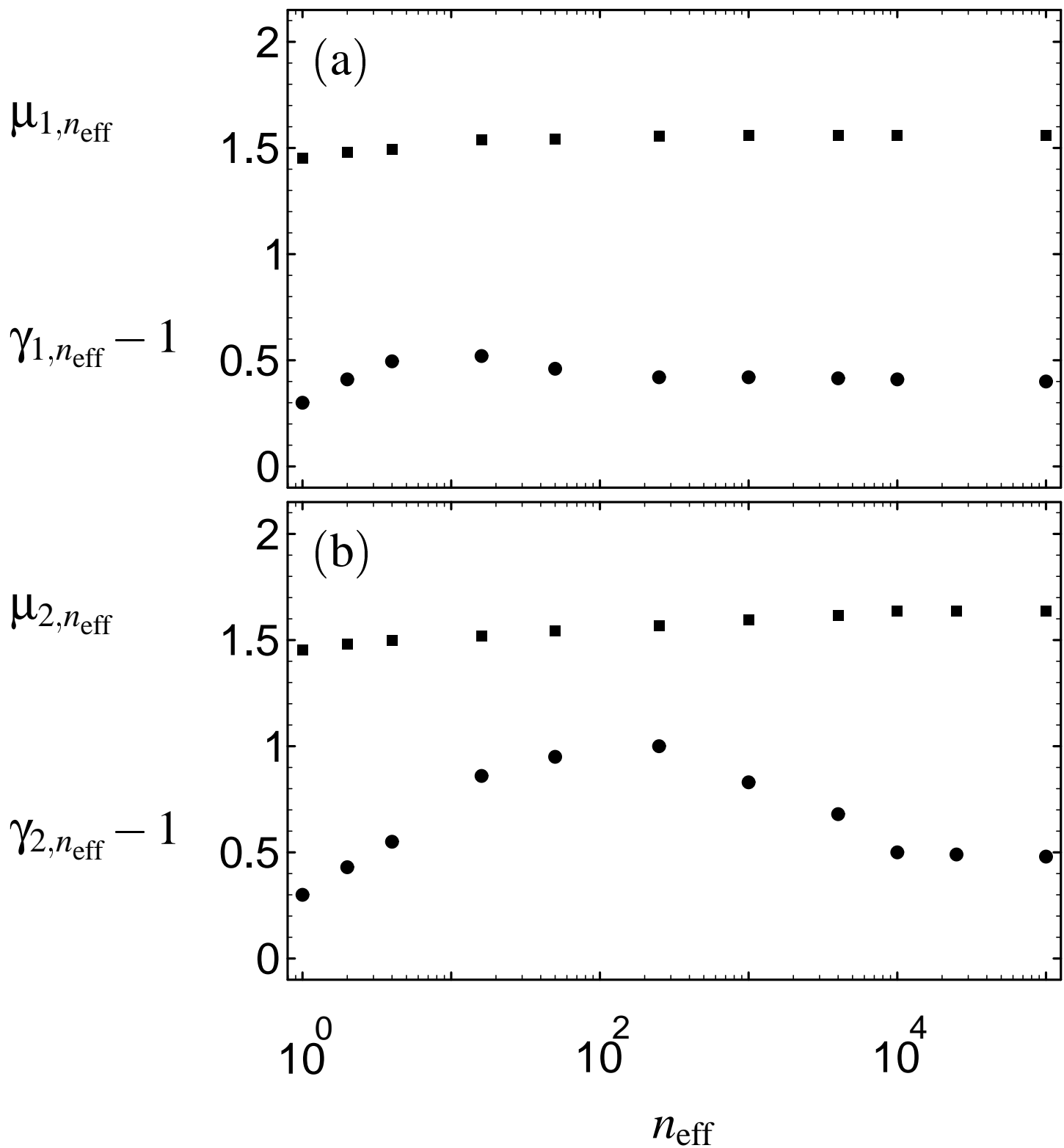
\end{document}